\documentclass[a4paper]{article}

\usepackage{INTERSPEECH2022}
\usepackage{hyperref}
\usepackage{url}
\usepackage{graphicx}
\usepackage{xcolor}
\usepackage{amsmath,amssymb}
\usepackage{hyperref}
\usepackage{algorithmic}
\usepackage{booktabs,eqparbox}
\usepackage{adjustbox}
\usepackage[linesnumbered, ruled]{algorithm2e}
\usepackage{tikz}
\usepackage{setspace}
\usepackage{multirow}
\usepackage{ctable}

\def\checkmark{\tikz\fill[scale=0.3](0,.35) -- (.25,0) -- (1,.7) -- (.25,.15) -- cycle;}
\def\crossmark{\text{\sffamily X}}

\title{A Comparative Study of Data Augmentation Techniques for Deep Learning Based Emotion Recognition}
\name{Ravi Shankar$^1$, Abdouh Harouna Kenfack$^2$, Arjun Somayazulu$^3$, Archana Venkataraman$^1$}
\address{
  $^1$Department of Electrical and Computer Engineering, Johns Hopkins University, USA\\
  $^2$Department of Applied Mathematics and Statistics, Johns Hopkins University, USA\\  
  $^3$Department of Computer Science, Johns Hopkins University, USA}
\email{rshanka3@jhu.edu, aharoun1@jhu.edu, asomaya1@jhu.edu, archana.venkataraman@jhu.edu}

\begin{document}

\maketitle
\begin{abstract}
Automated emotion recognition in speech is a long-standing problem.
While early work on emotion recognition relied on hand-crafted features and simple classifiers, the field has now embraced end-to-end feature learning and classification using deep neural networks.
In parallel to these models, researchers have proposed several data augmentation techniques to increase the size and variability of existing labeled datasets.
Despite many seminal contributions in the field, we still have a poor understanding of the interplay between the network architecture and the choice of data augmentation. 
Moreover, only a handful of studies demonstrate the generalizability of a particular model across multiple datasets, which is a prerequisite for robust real-world performance. 
In this paper, we conduct a comprehensive evaluation of popular deep learning approaches for emotion recognition. 
To eliminate bias, we fix the model architectures and optimization hyperparameters using the VESUS dataset and then use repeated 5-fold cross validation to evaluate the performance on the IEMOCAP and CREMA-D datasets. Our results demonstrate that long-range dependencies in the speech signal are critical for emotion recognition and that speed/rate augmentation offers the most robust performance gain across models.
\end{abstract}

\maketitle

\section{Introduction}
Human speech is a rich and varied medium that encapsulates not only semantic content, but also the speaker's mood and intent. 
While it is often easy for humans to decode the latter information, the same cannot be said for computers, particularly when it comes to parsing emotional cues from speech. 
At the same time, automated emotion recognition has the potential to greatly improve human-interactions~\cite{hci_emotion}.
For example, it can be used in call centers for on-the-fly sentiment analysis to provide better customer support~\cite{call_center_monitoring}.  
Likewise, smart car systems and forensic analysis of speech rely heavily on being able to understand the emotional state of the person~\cite{smart_vehicle_assistants, forensic_emotion} under study.
There are also medical applications for automated emotion recognition, such as diagnosing depression and anxiety from the speaking style, and automated monitoring of patients with bipolar disorder~\cite{smart_affective_emotion_recognition}.

Due to these and other applications, the area of emotion recognition has seen a steady growth in the past few years.
Deep neural networks, in particular, have allowed for a steady increase in performance on challenging acted and improvised emotional speech datasets~\cite{iemocap, cremaD, ravdess, vesus, savee}.
In the simplest case, hand-crafted features are fed into a deep network classifier to predict one of several emotional states. One such example is the work of~\cite{vesus}, in which the authors use a multi-layer perceptron to identify five different emotional states. 
This hand-crafted approach has largely been superceded by models that learn task-specific feature representations across time. 
Convolutional neural neural network (CNN) models operate on time-frequency inputs (e.g., Mel-frequency cepstral coefficients, chromagram) and use either 1-D or 2-D convolutions to learn the relevant filters for emotion classification~\cite{conv_emo_1, conv_emo_2, conv_emo_3, conv_emo_4}.
A 3-D convolutional model using the first and second order derivative of MFCCs has also been proposed in the literature by~\cite{acrnn}. 
Recurrent neural networks (RNNs) for emotion recognition~\cite{rnn_emo_1, rnn_emo_2, rnn_emo_3} are also popular due to their ability to process temporal sequences. 
In fact, an attention based Bi-LSTM model proposed by~\cite{bilstm_attention} simultaneously learns not only the emotional label but which segments of an utterance are responsible for the corresponding prediction. 
Finally, the transformer architecture is often used for multi-modal emotion recognition, where the inputs are audio and text~\cite{transformer_seq2seq} or audio and video~\cite{transformer_emotion_recognition}. Transformers and RNNs have also been combined with other deep network modules, such as CNNs, for improved emotion recognition.

The flip-side to the popularity of deep learning models is the need for larger and more varied training data. To mitigate this need, researchers have proposed data augmentation strategies that can expand the training dataset by few orders of magnitude.
The most common data augmentation techniques for emotion recognition are to add random noise~\cite{noise_aug_emotion}, to vary the sampling rate of the input speech signal~\cite{speed_perturbation}, to manipulate the spectral content of the utterances~\cite{spec_augment}, and a mixing between these approaches~\cite{mixup_emotion}.
Beyond signal manipulations, a recently proposed augmentation strategy known as copy-paste~\cite{copy_paste} randomly combines neutral and emotional utterances.
This strategy allows the models to extract time-limited emotional cues that may not be present for the entire utterance duration. 
While all of these augmentation methods have led to a significant increase in emotion recognition accuracy, they are tested on a specific model and experimental dataset. 
Therefore, we have little understanding about which models and augmentation strategies provide robust and generalizable performance gains.

\begin{figure*}[!t]
  \centering
  \includegraphics[width=0.85\textwidth, height=5.5cm]{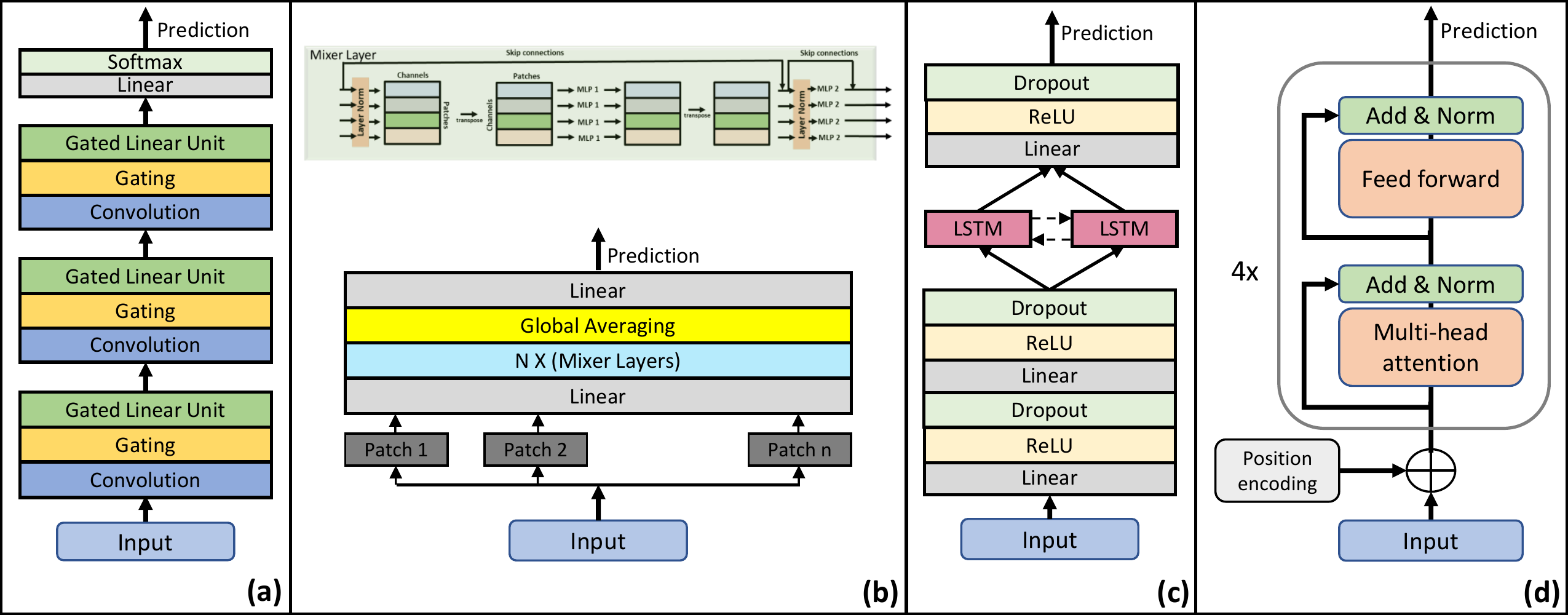}
  \caption{Neural network architectures used for emotion classification: (a) Gated CNN, (b) MLP mixer, (c) Bi-LSTM with attention, and (d) transformer with multi-head attention. We identify the best model architecture via cross validation on the VESUS dataset. This strategy mitigates bias and over-fitting during our final evaluation on IEMOCAP and CREMA-D.}
  \label{fig:neural_network_architectures}
  \vspace{-4mm}
\end{figure*}

In this paper, we conduct a comprehensive evaluation of canonical deep network architectures and data augmentation strategies. Our architectures consist of (1) a convolutional model, (2) a fully-connected model similar to MLP, (3) a recurrent model, and (4) a transformer. For each model, we evaluate five augmentation strategies: original/no augmentation, adding noise at different SNRs, re-sampling the timescales of the training utterances, masking time/frequency blocks in the spectrogram, and concatenating neutral and emotional utterances to mitigate class imbalance. To avoid over-fitting, we fix the network architectures and hyperparameters based on the  VESUS dataset~\cite{vesus}. We then perform a repeated $k$-fold cross validation of each (model, augmentation) pair on the more challenging IEMOCAP~\cite{iemocap} and CREMA-D~\cite{cremaD} datasets.

\section{Methods}
In order to provide a comprehensive evaluation of the interplay between network architecture and data augmentation, we select one model from each of the following classes: (1) convolutional network, (2) fully-connected network, (3) recurrent network, and (4) transformer. We also query four distinct data augmentation strategies used in speech analysis tasks. 

\vspace{-1mm}
\subsection{Network Architectures}

This section details the selected architectures. These networks are adapted from existing work in language modeling, speech recognition, and multimodal audio/video processing. The input to these models are the MFCCs feature vectors computed over a window of 5ms with the same temporal shift. These representations are commonly used for emotion recognition~\cite{speech_emo_features}.

\vspace{-2mm}
\subsubsection{Gated Convolutional Network}
The gated convolutional neural network (Gated-CNN) was proposed by~\cite{gated_conv} for language modeling using short contexts. 
The Gated-CNN extends a typical CNN by constructing two parallel convolution operations for each input, with the output of one branch passed through a sigmoid activation, which acts as a gating. The two branches are combined element-wise. Mathematically, let $\mathbf{S} \in \mathbb{R}^{T \times F}$ be the input signal, and $\mathbf{w_s}$ and $\mathbf{w_g}$ denote the convolutional kernels. The Gated-CNN output is:
\vspace{-1mm}
\begin{equation} \label{eqn:GatedCNN}
    \mathbf{Y} = LN((\mathbf{S} * \mathbf{w_s}) \odot \sigma(\mathbf{S} * \mathbf{w_g})),
\end{equation}
where $\odot$ is the element-wise Hadamard product, and $LN$ denotes layer normalization. At a high level, the sigmoid gating serves as an information regulator for the downstream task.

As seen in Fig.~\ref{fig:neural_network_architectures}(a), we feed the output of the CNN cascade into one fully-connected layer, followed by a softmax operation to classify the correct emotion category. 

\vspace{-2mm}
\subsubsection{MLP-Mixer}
The multilayer perceptron mixer (MLP-mixer) is an innovative network architecture for processing 2D inputs. It was first proposed in the context of image processing~\cite{mlp_mixer}, but here we adapt it to time-frequency representations of speech. 

Our MLP-mixer consists of a stack of sequential time and frequency mixing layers interleaved with layer normalization to address the vanishing gradient and covariant shift issue (Fig.~\ref{fig:neural_network_architectures}(f)). 
This operation facilitates information sharing across time and frequency bands in a computationally efficient manner.

Mathematically, let $\mathbf{S} \; \epsilon \; \mathbb{R}^{T \times F}$ be the input time-frequency representation, where $T$ is the number of frames in the speech signal, and $F$ is the resolution of the MFCCs. The output at the first layer of MLP-mixer can be written as follows: 
\vspace{-1mm}
\begin{align} \label{eqn:MLP1}
    \mathbf{U}_f &= \mathbf{S}_{*,f} + \mathbf{W}_2 \times \sigma (\mathbf{W}_1 \times LN(\mathbf{S})_{*,f}) \\
    \mathbf{V}_t &= \mathbf{U}_{t,*} + \mathbf{W}_3 \times \sigma (\mathbf{W}_4 \times LN(\mathbf{U})_{t,*})
\end{align}
for $f=1,\ldots,F$ and $t=1,\ldots,T$. Once again, $LN$ denotes the layer normalization operation, and the variables $\{\mathbf{W}_i\}$ are the weight matrices. Subsequent layers of the MLP-mixer follow likewise. As seen in Fig.~\ref{fig:neural_network_architectures}(b), we use global average pooling and  a linear layer for the final classification task.  

\vspace{-2mm}
\subsubsection{Bi-LSTM with Attention}
Our recurrent module consists of the Bi-LSTM with attention model proposed by~\cite{bilstm_attention}. The input is a sequence of frequency-based feature vectors, such as the frame-wise spectrogram or MFCCs. In parallel to processing the inputs, an attention value for each time point is obtained by computing a dot-product with a learnable vector. At a high level, the attention captures how important a particular time point is for the downstream task. A final feed-forward layer combines the data representation and attention for the emotion classification (Fig.~\ref{fig:neural_network_architectures}(d)). 

\begin{table*}[h]
    \centering
    \begin{tabular}{|c|c|c|c|c|c|c|}
    \hline
    Model & $\#$Layers & $\#$Nodes(final) & $\#$Kernels & $\#$Attention Heads & $\#$LSTM cells & Patch size \\
    \hline
    Gated-CNN & $\checkmark$ & $\checkmark$ & $\checkmark$ & $\crossmark$ & $\crossmark$ & $\crossmark$ \\
    \hline
    MLP-mixer & $\checkmark$ & $\checkmark$ & $\crossmark$ & $\crossmark$ & $\crossmark$ & $\checkmark$ \\
    \hline
    Bi-LSTM & $\checkmark$ & $\checkmark$ & $\crossmark$ & $\crossmark$ & $\checkmark$ & $\crossmark$ \\
    \hline
    Transformer & $\checkmark$ & $\checkmark$ & $\crossmark$ & $\checkmark$ & $\crossmark$ & $\crossmark$ \\
    \hline
    \end{tabular}
    \caption{Hyperparameters of the network architectures optimized using the VESUS dataset.}
    \label{tab:hyperparam_opt}
    \vspace{-6mm}
\end{table*}

\vspace{-2mm}
\subsubsection{Transformer Network}
Transformers have been also proposed for language and emotion modeling over longer contexts~\cite{transformer_seq2seq, transformer_emo}. 
The central idea of this architecture is to exploit short and long-term dependencies via a dot product operation on the input sequence.
This is combined with multiple self-attention operations, which act as feature extractors for the downstream emotion classification task. Mathematically, the self-attention can be written as follows:
\vspace{-1mm}
\begin{equation}\label{eqn:attn}
    \mathbf{A}(\mathbf{Q}, \mathbf{K}, \mathbf{V}) = softmax \left(\frac{\mathbf{Q}\mathbf{K}^T}{\sqrt{d_k}}\right)\mathbf{V}    
\end{equation}
We use the linear transformations of the input sequence to construct the \textit{query} $\mathbf{Q}$, the \textit{key} $\mathbf{K}$, and the \textit{value} $\mathbf{V}$ in Eq.~(\ref{eqn:attn}).

We use sinusoidal position embedding to identify the relative positioning of a single frame in the input sequence. Further, each attention layer has a residual connection that combines the input directly with the output of the attention mechanism to facilitate gradient flow during backpropagation (Fig.~\ref{fig:neural_network_architectures}(b)).

\begin{figure}[t]
  \centering
  \includegraphics[width=0.95\linewidth, height=4.3cm]{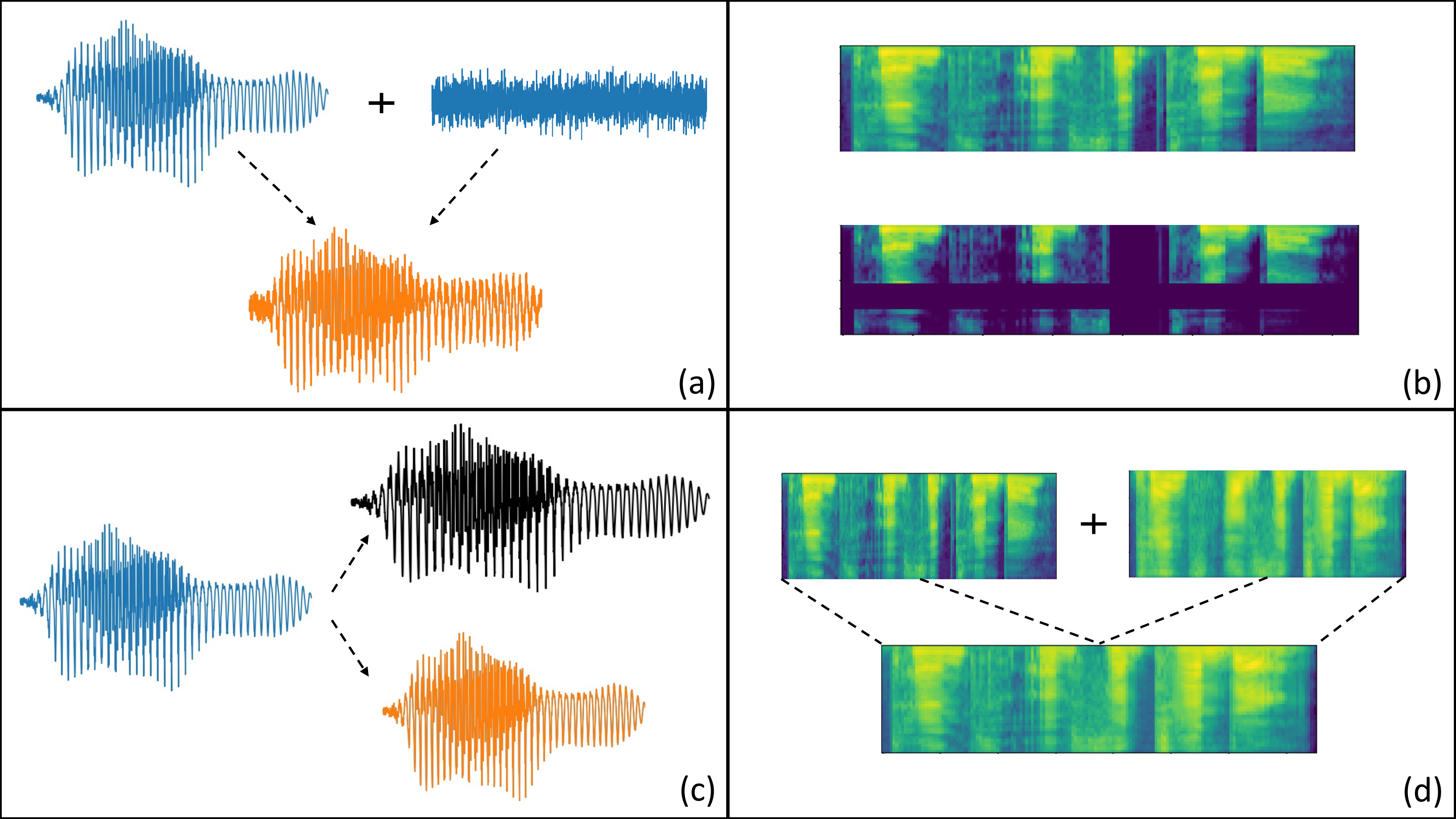}
  \caption{Augmentation schemes used for training emotion classification models. (a) Additive Noise (b) Spectral Augmentation (c) Speed Perturbation and (d) Copy-Paste Augmentation}
  \label{fig:augmentation_schemes}
  \vspace{-5mm}
\end{figure}

\subsection{Data Augmentation Strategies}
As seen in Fig.~\ref{fig:augmentation_schemes}, we evaluate four data augmentation strategies that have been recently proposed for speech analysis.

\smallskip \noindent \textbf{1) Additive Noise:} We add white Gaussian noise to the original audio signal at a specified signal-to-noise ratio (SNR). In this work, we consider three SNR levels of 10dB, 20dB and 30dB.

\smallskip \noindent \textbf{2) Spectral Augmentation (SpecAug):} This data augmentation strategy was proposed by~\cite{spec_augment} and has led to significant improvement in speech recognition. 
The algorithm involves masking a random interval of the spectrogram along the frequency axis.  
This masking is followed by a random control point registration of the time-frequency representation (see Fig.\ref{fig:augmentation_schemes}(b)). 
In this work, we also run SpecAug for time/frequency masking and for time-frequency masking to compare their relative effects.

\smallskip \noindent \textbf{3) Speed Perturbation:} Introduced for speech recognition \cite{speed_perturbation}, this augmentation strategy changes the rhythm of the signal will preserving the original pitch.
In this work, we augment the original input data with samples that have been slowed by 10\% (i.e., speed perturbation of 0.9) and samples that have been sped up by 10\% (i.e., speed perturbation of 1.1).
We also evaluate the effect of mixing both perturbation types. 

\smallskip \noindent \textbf{4) Copy-Paste:} This algorithm was proposed by~\cite{copy_paste} to increase the diversity of training examples in the dataset for emotion recognition. 
The key idea in Copy-Paste technique is to concatenate neutral and emotional speech but maintain the emotion label for the augmented utterance (Fig.\ref{fig:augmentation_schemes}(d)).
This technique capitalizes on the fact that emotions are expressed in a segmental manner, meaning that only portions of the utterance carry emotion-specific attributes.
In this work, we use Copy-Paste to balance the training examples across all classes. 

\begin{table*}[h]
    \centering
    \bgroup
    \renewcommand{\arraystretch}{1.2}
    \begin{adjustbox}{width=\textwidth}
    \begin{tabular}{c !{\vrule width 1pt} c c c c c c c c c c c}
    
    \specialrule{1pt}{0pt}{0pt}
    Model & NoAug & Noise(10db) & Noise(20db) & Noise(30db) & SpecAug(TF) & SpecAug(T) & SpecAug(F) & Speed(0.9) & Speed(1.1) & Speed(M) & CopyPaste \\
    \specialrule{1pt}{0pt}{0pt}
    
    Gated-CNN & $0.56\pm.02$ & $0.54\pm.03$ & $0.53\pm.02$ & $0.53\pm.04$ & $0.54\pm.03$ & $0.54\pm.003$ & $0.56\pm0.01$ & $0.56\pm.05$ & $0.56\pm.02$ & $\underline{0.59\pm.01}$ & $\mathbf{0.6\pm0.01}$ \\

    MLP-mixer & $0.52\pm.02$ & $0.51\pm.01$ & $0.54\pm.02$ & $0.5\pm.01$ & $0.54\pm.01$ & $0.51\pm.03$ & $0.53\pm.03$ & $0.53\pm.03$ & $\mathbf{0.57\pm.01}$ & $0.51\pm.02$ & $0.52\pm.03$ \\

    Bi-LSTM & $0.52\pm.02$ & $0.51\pm.02$ & $0.52\pm.02$ & $0.52\pm.02$ & $0.54\pm.02$ & $0.55\pm.001$ & $0.54\pm.01$ & $\mathbf{0.57\pm.02}$ & $0.55\pm.01$ & $\underline{0.56\pm.01}$ & $\underline{0.56\pm.01}$ \\

    Transformer & $\mathbf{0.59\pm.02}$ & $0.52\pm.02$ & $0.52\pm.04$ & $0.54\pm.01$ & $0.56\pm.01$ & $0.54\pm.02$ & $0.55\pm.02$ & $0.54\pm.02$ & $\underline{0.58\pm.03}$ & $0.55\pm.03$ & $0.54\pm.05$ \\
    \specialrule{1pt}{0pt}{0pt}
    
    Gated-CNN & $0.54\pm.02$ & $0.51\pm.03$ & $0.51\pm.02$ & $0.51\pm.04$ & $0.51\pm.04$ & $0.52\pm.01$ & $0.54\pm.01$ & $0.53\pm.05$ & $0.54\pm.02$ & $0.57\pm.01$ & $\mathbf{0.59\pm.005}$ \\
    
    MLP-mixer & $0.5\pm.02$ & $0.49\pm.01$ & $0.51\pm.02$ & $0.47\pm.02$ & $0.51\pm.01$ & $0.49\pm.04$ & $0.50\pm.04$ & $0.51\pm.03$ & $\mathbf{0.54\pm.01}$ & $0.50\pm.02$ & $0.49\pm.02$ \\
    
    Bi-LSTM & $0.50\pm.02$ & $0.5\pm.03$ & $0.5\pm0.02$ & $0.51\pm0.02$ & $0.53\pm.01$ & $0.53\pm.01$ & $0.53\pm.01$ & $\mathbf{0.56\pm.02}$ & $0.53\pm.002$ & $0.56\pm.01$ & $0.54\pm.01$ \\
    
    Transformer & $\mathbf{0.56\pm.03}$ & $0.51\pm.02$ & $0.50\pm.04$ & $0.52\pm.01$ & $\underline{0.55\pm.01}$ & $0.52\pm.01$ & $0.53\pm.02$ & $0.53\pm.02$ & $0.56\pm.03$ & $0.54\pm.03$ & $0.52\pm.04$ \\
    \specialrule{1pt}{0pt}{0pt}
    \end{tabular}
    \end{adjustbox}
    \egroup
    \caption{Emotion recognition using the IEMOCAP dataset. \textbf{Top:} unweighted average accuracy. \textbf{Bottom:} weighted F1 score}
    \vspace{-5mm}
    \label{tab:IEMOCAP}
\end{table*}

\begin{table*}[h]
    \centering
    \bgroup
    \renewcommand{\arraystretch}{1.2}
    \begin{adjustbox}{width=\textwidth}
    \begin{tabular}{c !{\vrule width 1pt} c c c c c c c c c c c}
    \specialrule{1pt}{0pt}{0pt}
    Model & NoAug & Noise(10db) & Noise(20db) & Noise(30db) & SpecAug(TF) & SpecAug(T) & SpecAug(F) & Speed(0.9) & Speed(1.1) & Speed(M) & CopyPaste \\
    \specialrule{1pt}{0pt}{0pt}
    
    Gated-CNN & $0.64\pm.01$ & $0.64\pm.004$ & $0.64\pm.01$ & $0.64\pm.003$ & $0.64\pm.01$ & $0.64\pm.01$ & $0.64\pm0.01$ & $\underline{0.65\pm.01}$ & $0.64\pm.004$ & $\mathbf{0.66\pm.003}$ & $0.65\pm0.01$ \\
    
    MLP-mixer & $0.53\pm.004$ & $0.50\pm.01$ & $0.49\pm.01$ & $0.49\pm.01$ & $\underline{0.54\pm.01}$ & $0.53\pm.02$ & $0.53\pm.01$ & $\underline{0.54\pm.002}$ & $0.53\pm.004$ & $\mathbf{0.55\pm.003}$ & $0.52\pm.01$ \\
    
    Bi-LSTM & $0.61\pm.004$ & $0.63\pm.003$ & $0.63\pm.004$ & $0.62\pm.01$ & $0.61\pm.003$ & $0.61\pm.01$ & $0.62\pm.002$ & $\underline{0.65\pm.004}$ & $0.64\pm.01$ & $\mathbf{0.66\pm.01}$ & $0.62\pm.01$ \\
    
    Transformer & $0.62\pm.003$ & $0.61\pm.004$ & $0.62\pm.004$ & $0.62\pm.01$ & $0.62\pm.01$ & $0.63\pm.01$ & $0.62\pm.01$ & $\underline{0.64\pm.001}$ & $0.63\pm.002$ & $\mathbf{0.65\pm.004}$ & $0.63\pm.01$ \\
    \specialrule{1pt}{0pt}{0pt}
    
    Gated-CNN & $0.64\pm.01$ & $0.64\pm.004$ & $0.64\pm.01$ & $0.64\pm.003$ & $0.64\pm.01$ & $0.64\pm.01$ & $0.64\pm.01$ & $\underline{0.65\pm.01}$ & $0.64\pm.004$ & $\mathbf{0.66\pm.003}$ & $0.64\pm.01$ \\
    
    MLP-mixer & $0.52\pm.004$ & $0.50\pm.01$ & $0.49\pm.01$ & $0.49\pm.01$ & $0.53\pm.01$ & $0.52\pm.01$ & $0.53\pm.003$ & $0.53\pm.003$ & $0.53\pm.004$ & $\mathbf{0.55\pm.002}$ & $0.51\pm.004$ \\
    
    Bi-LSTM & $0.60\pm.01$ & $0.62\pm.003$ & $0.63\pm0.004$ & $0.62\pm0.01$ & $0.60\pm.004$ & $0.61\pm.01$ & $0.62\pm.001$ & $\underline{0.65\pm.004}$ & $0.64\pm.01$ & $\mathbf{0.66\pm.01}$ & $0.62\pm.01$ \\
    
    Transformer & $0.62\pm.003$ & $0.60\pm.003$ & $0.62\pm.01$ & $0.61\pm.01$ & $0.62\pm.01$ & $0.62\pm.01$ & $0.62\pm.01$ & $\underline{0.63\pm.001}$ & $\underline{0.63\pm.002}$ & $\mathbf{0.64\pm.01}$ & $\underline{0.63\pm.01}$ \\
    \specialrule{1pt}{0pt}{0pt}
    \end{tabular}
    \end{adjustbox}
    \egroup
    \caption{Emotion recognition using the CREMA-D dataset. \textbf{Top:} unweighted average accuracy. \textbf{Bottom:} weighted F1 score}
    \vspace{-7mm}
    \label{tab:CREMAD}
\end{table*}

\subsection{Implementation Details}
The input to each model consists of the standard 23-dimensional Mel filterbank features extracted from the speech utterance using a window of size 5ms and the same shift. 
We fix the hyperparameters for each architecture (e.g., depth, nodes, \#filters, \#attention heads, etc) using the VESUS dataset~\cite{vesus} to eliminate bias when evaluated on the testing datasets, IEMOCAP~\cite{iemocap} and CREMA-D~\cite{cremaD}. 
Table~\ref{tab:hyperparam_opt} summarizes the hyperparameters optimized for each model.
In addition, to account for the effects of random initialization, we train/test each model multiple times to produce a confidence interval for the emotion recognition. 
We train each model on a single A32 GPU in PyTorch~\cite{pytorch}. 
The Gated-CNN, MLP mixer and Bi-LSTM are optimized using Adam optimizer with a fixed learning rate of 1e-3. 
The Transformer model starts from the same learning rate but has a step-wise learning rate decay schedule of 0.95 per step. 
\vspace{-1mm}
\section{Experimental Results}

\subsection{Emotional Speech Datasets}
As noted above, we select the architectures and hyperparameters of each model via five-fold cross validation on the VESUS dataset~\cite{vesus}. 
VESUS contains a total of 10 actors (5 males and 5 females) reading a script in multiple emotion categories.
We create the five folds by pairing one male and one female actor. 
VESUS also contains crowd-sourced emotional ratings for each utterance. For robustness, we use only utterances that have been correctly labeled as emotional by at least half of the raters.

Our evaluation datasets in this work are IEMOCAP~\cite{iemocap} and CREMA-D~\cite{cremaD}.
IEMOCAP contains emotional speech collected by from conversations between a male and a female actor. 
The dataset contains 10 actors in total, grouped into 5 sessions.
Unlike VESUS, it also contains both scripted and improvised utterances.
To remain consistent with the literature, we select four primary emotions (neutral, angry, sad, and happy) for our classification task.  
We conduct 5 fold session-independent experiments by training each model on 4 sessions and testing on the held-out session, cyclically. 
CREMA-D is a multimodal corpus with audio-video recording by 91 speakers. 
We select audio files corresponding to the four primary emotions used in the IEMOCAP experiment and randomly partition the speakers in five groups to perform a five-fold cross validation. 
\vspace{-1mm}
\subsection{Emotion Recognition Performance}
Table~\ref{tab:IEMOCAP} summarizes the performance of each model and augmentation strategy on the IEMOCAP dataset. For convenience, we have bolded the best performance and underlined a close second place (e.g., within a standard deviation). 
As seen, the best model achieves a maximum accuracy score of 0.6 using the Gated-CNN architecture with the copy-paste augmentation strategy. 
The corresponding F1 score is high indicating a roughly balanced result across emotion classes. 
At a high level, convolutions facilitate the propagation of local information across multiple layers.
This works well with the copy-paste algorithm, as only portions of the combined utterances are emotional.
The Bi-LSTM and MLP-mixer models struggle to capture these local characteristics, in part due to the dyadic setup of IEMOCAP, where emotions are sporadic in nature. 
The transformer comes close to the best performance without any data augmentation, which suggests that the self-attention helps to improve the sensitivity for certain classes. 
Focusing on the augmentation strategies, we note that additive noise and SpecAug are poor augmentation schemes for IEMOCAP, possibly due to the varied semantic and emotional nature of the recording sessions. 
On the other hand, speed perturbation provide a robust increase in performance across all model classes.

\begin{figure}[t]
  \centering
  \includegraphics[width=0.72\linewidth, height=5.2cm]{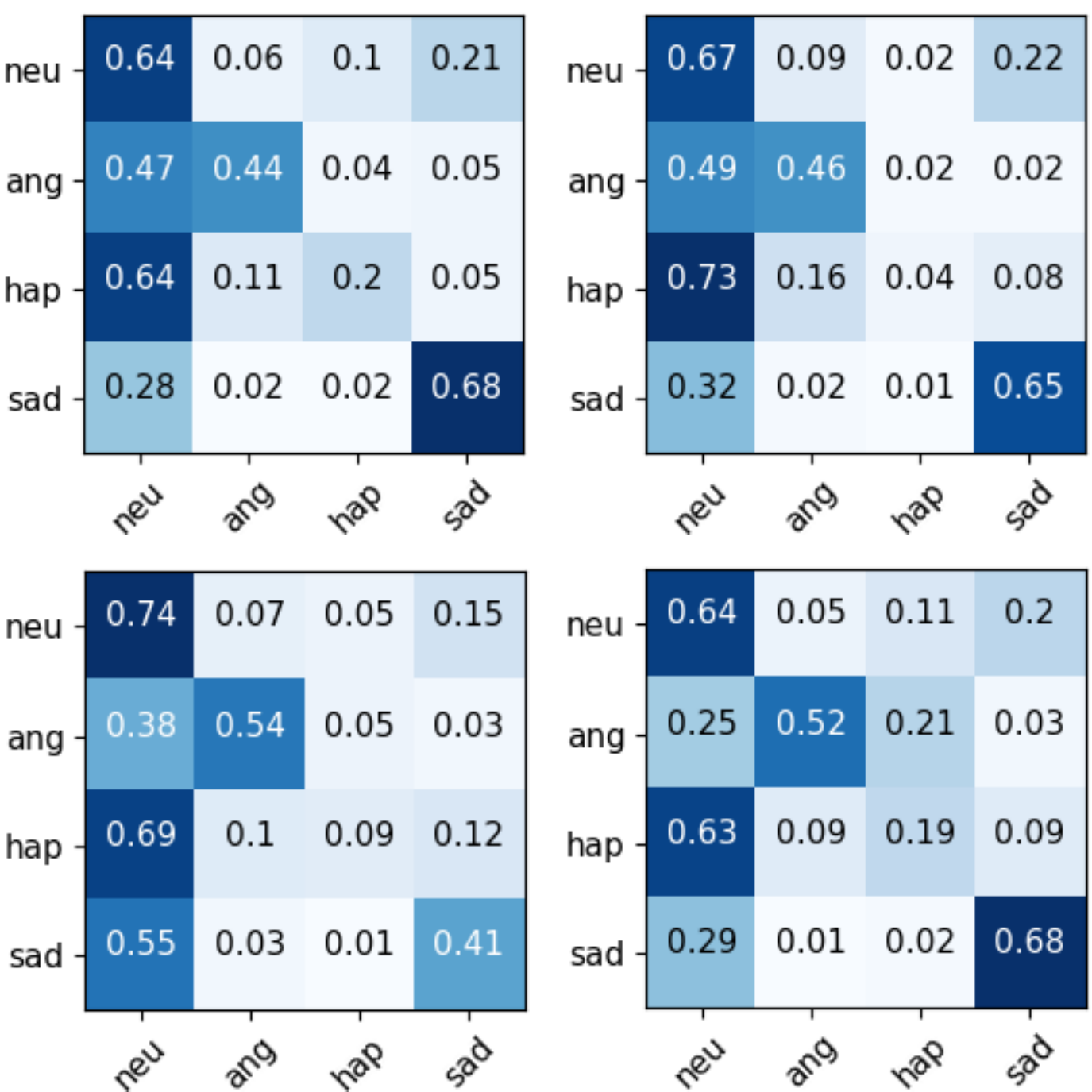}
  \caption{IEMOCAP Confusion matrices. Clockwise from top left: Gated Conv, MLP mixer, BiLSTM and Transformer.}
  \label{fig:IEMOCAP_confusion}
  \vspace{-6mm}
\end{figure}

\begin{figure}[t]
  \centering
  \includegraphics[width=0.72\linewidth, height=5.2cm]{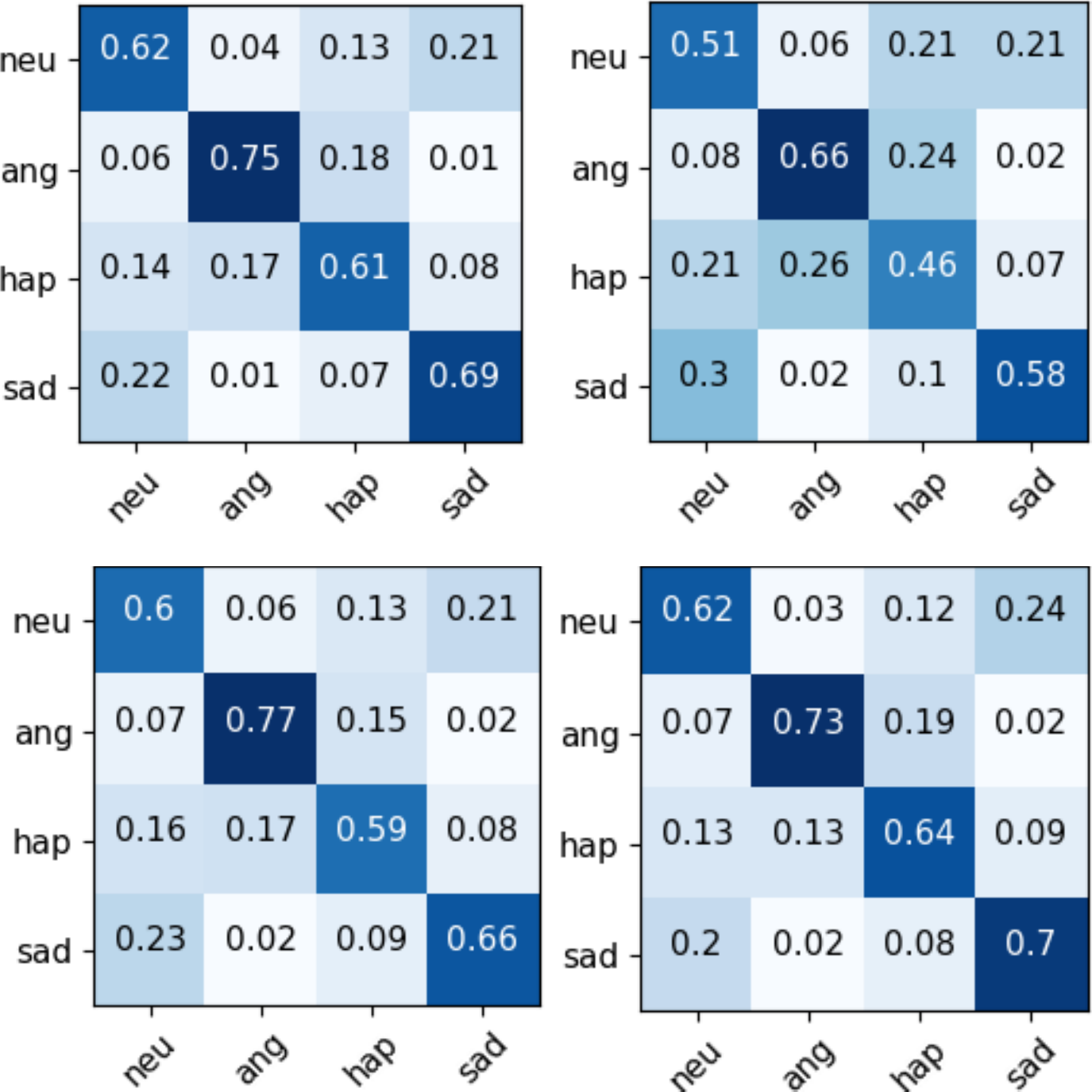}
  \caption{CREMA-D Confusion matrices. Clockwise from top left: Gated Conv, MLP mixer, BiLSTM and Transformer.}
  \label{fig:CREMAD_confusion}
  \vspace{-6mm}
\end{figure}

Table~\ref{tab:CREMAD} reports the performance of each model and augmentation strategy on the CREMA-D dataset.
First, we note that the overall performance is significantly higher than IEMOCAP.
This is because CREMA-D is an acted emotional dataset collected in a noise free environment, whereas IEMOCAP contains ``messy" dyadic interactions. 
Once again, we note that the Gated-CNN model performs the best, followed by the transformer. 
Across model architectures, we see that the MLP mixer has the worst performance, as it does not employ any type of temporal continuity or smoothness constraint in its formulation. 
With regards to the augmentation, the speed perturbation is the clear winner across model classes.
Once again, noise augmentation does not improve the metrics. 
Masking of time/frequency bands is similar to no augmentation meaning that most deep neural networks can fill in the information gap, i.e, interpolate in the masked region of training examples. 
Taken together with the IEMOCAP results, we conclude that having sufficient variability in the speech rhythm is crucial for emotion recognition. 

Finally, Fig.~\ref{fig:IEMOCAP_confusion} and Fig.~\ref{fig:CREMAD_confusion} illustrate the confusion matrices for the best performing augmentation strategy by each model on IEMOCAP and CREMA-D. On IEMOCAP, happy is the worst performing class, likely due to the smaller number of samples. Angry and sad are often confused with neutral, perhaps due to variations in the utterance lengths and emotional expressions across actors. 
CREMA-D has a balanced number of samples across emotion categories, which leads to higher sensitivity. This can be seen in the diagonally-dominant confusion matrices across model classes. Once again, happy is the most challenging class, which suggests that future studies should focus on data augmentation schemes targeted at this class.



\vspace{-2mm}
\section{Conclusions}
This paper presents a comprehensive study of the interplay between model architectures and data augmentation in the context of speech emotion recognition. 
We evaluate a convolutional, fully-connected, recurrent and transformer architecture on two distinct datasets. 
We optimize the model hyper-parameters on VESUS and then evaluated them on IEMOCAP and CREMA-D for an unbiased comparison. 
We conclude that speed perturbation is a robust augmentation strategy across network architectures and datasets.
The copy-paste augmentation is also useful for convolutional models.
Ultimately, this study can be used as a helpful guide for future researcher when designing and implementing neural architectures for emotional speech recognition. 

\section{Acknowledgements}
We thank Deeksha M. Shama (deeksha1@jhu.edu) and Natalie Aw (naw1@jhu.edu) for their helpful suggestions in designing experiments and setting it up on computing cluster. 

\bibliographystyle{IEEEtran}

\bibliography{main.bbl}

\end{document}